  \documentclass[12pt,preprint]{aastex}
 \usepackage{ifpdf}
 \usepackage{epsfig}
  \usepackage[latin1]{inputenc}
 \usepackage{amsfonts}
 \usepackage{amsthm}
 \usepackage{amssymb, latexsym, amsmath}
 \usepackage{amsbsy}
\usepackage{amstext}
\usepackage{amsopn}
\usepackage{amsgen}
\usepackage[dvips]{color}
 \usepackage{graphicx}
  \usepackage{color}
\newcommand{\eb}{\begin{equation}}
\newcommand{\ee}{\end{equation}}
\newcommand{\st}{KIC 8462852~}
\shorttitle{Photometry and astrometry of KIC 8462852}
\shortauthors{Makarov \& Goldin}
\begin{document}
\title{Photometric and astrometric vagaries of the enigma star KIC 8462852}
\author{Valeri V. Makarov}
\email{vvm@usno.navy.mil} 
\affil{US Naval Observatory, 3450 Massachusetts Ave NW, Washington DC 20392-5420, USA}
\author{Alexey Goldin}
\email{alexey.goldin@gmail.com} 
\affil{Teza Technology, 150 N Michigan Ave, Chicago IL 60601, USA}

\date{Accepted . Received ; in original form }

%%\pagerange{\pageref{firstpage}--\pageref{lastpage}} \pubyear{2015}
%\maketitle
\label{firstpage}
\begin{abstract}
We apply a PCA-based pre-whitening method to the entire collection of main Kepler mission long-cadence data
for KIC 8462852 spanning four years. This technique removes the correlated variations of instrumental origin in both the detected 
light curves and astrometry, resolving intrinsic changes in flux and image position of less than 100 ppm and 1 mas,
respectively. Beside the major dips in the light curve during mission quarters 8 and 16, when the flux dropped by up to 20\%, we confirm multiple smaller dips across the time span of observation with
amplitudes ranging from 0.1\% to 7\%. A variation of flux with a period of 0.88 d and a half-amplitude of approximately 90 ppm is confirmed in the PCA-cleaned data. We find that the phase of the wave is steady over the entire 15-month interval. We confidently detect a weak variability-induced
motion (VIM) effect in the cleaned astrometric trajectories, when the moment-based centroids shift synchronously with
the flux dips by up to 0.0008 pixels on the detector. The inconsistent
magnitude and direction of VIM effects within the same quarter point at more than one source of
photometric variability in the blended image. The 0.88 d periodicity comes from a different source, not from the
target star KIC 8462852. We discuss
a possible interpretation of the bizarre properties of the source as a swarm of interstellar junk (comets and planetoids) crossing the line of sight to the star and its optical companions at approximately 7 mas per year.
\end{abstract}

%\begin{keywords}
%binaries: celestial mechanics -- Moon -- planets and satellites: dynamical evolution and stability.
%\end{keywords}

\section{Introduction}
The star \st = TYC 3162-665-1, with unique and strange photometric features in the Kepler long-cadence light curves, was
serendipitously discovered by the Planet Hunters team \citep{boy} who routinely inspected thousands of targets
with candidate exoplanet transit events. The star underwent a series of dimming events during the 4
years of nominal Kepler mission of variable depth and duration, ranging from $\sim 1000$ parts per million (ppm) in
relative flux to $\sim 210\,000$ ppm. The most unusual features of these events are the apparent lack of
periodicity and repeatability and asymmetry of their shape. The discoverers collected follow-up observations of the star, including spectroscopic and radial velocity measurements and high-resolution imaging and speckle interferometry. These additional data
allowed the authors to rule out a number of hypotheses, such as a double star, a very young star with stochastic
variability, and an evolved supergiant of the R Coronae Borealis type. A system of large comets on highly eccentric orbits
or dust and rock debris resulting from a catastrophic collision of planets emerged as the likeliest explanation.
However, neither infrared \citep{mar} nor submillimetre \citep{tho} continuum fluxes show any significant excess,
practically disproving the presence of dusty material around the star. Somewhat heuristically, artificial manipulation
of the stellar irradiation by an extraterrestrial intelligence has been proposed, and efforts to detect other
signals or signs of such an activity have been made \citep{abe}. The characteristics of the detected photometric
events may be consistent with occultations by giant artificial megastructures \citep{wri}.

In this paper, we revisit the data collected by the Kepler main mission for \st and for other stars observed on the
same CCD detectors as this star. Our aim is to contribute to the in-depth analysis of Kepler data with the following updates in
the approach.
\begin{itemize}
\item
We use the raw Simple Aperture Photometry (SAP) light curves as opposed to the Pre-search Data Conditioning (PDC)
fluxes derived by the data processing pipeline \citep{chr}. The de-trending procedures that fit low-order polynomials
to remove the slow variations of flux in the raw data are not always sufficient to mitigate the common instrumental
perturbations shared by targets on the same channel, and can in fact distort the light curve on time scales longer
than a few days. Be mindful that the main purpose of PDC was to provide the most reliable information about
possible exoplanet transits, likely at some expense to longer-term variability.
\item
To remove most of the correlated perturbation of instrumental origin common to all targets within a given CCD
channel, we design and apply an ad hoc pre-whitening technique based on Principal Component Analysis (PCA).
The method is described in Sect. \ref{pca.sec}. This allows us to come as close as possible to
the intrinsic variability of the star at a level of 20 ppm and for time scales up to 90 days
after removing the much greater detrimental variations of instrumental origin.
\item
For the first time, we analyze in detail the astrometric data gathered by Kepler, i.e., the photocentre centroids
measured at the same cadence and simultaneously with the flux. Synchronous astrometric excursions of centroids, called Variability
Induced Motion (VIM),
provide a powerful method of detecting double stars or other events of signal blending \citep{mak}. 
VIM occurs when one of the blended sources in the aperture is
photometrically variable. Since the recorded centroid corresponds to the first moment of the distribution of
flux over the pixel grid, which is sensitive to the flux ratio of the blended components, the photocenter moves
in a strongly correlated way with the total flux. The collection of Kepler long cadence data is full of VIM
manifestations, many of the intended targets being blended with something else. Correlated
excursions of instrumental origin due to pointing events, micrometeoroid strikes, etc., are efficiently removed or suppressed by the same
pre-whitening technique with appropriate modifications. This allows us to detect genuine astrometric signals
at a level $\sim 0.0002$ pixels, which corresponds to $\sim 1$ mas on the sky. A detailed study of the detected VIM 
signals for the star in question is given in Section \ref{eve.sec}.
\item
Using the least-squares Lomb periodogram, we analyze the sinusoidal signals in both the pre-whitened light curves and astrometric
trajectories (Section \ref{per.sec}). Using the data in quarters 9 through 13 when the star was the most quiescent
the stability of the phase of the 0.88 d variation is evaluated.
\end{itemize}
Given the previously accumulated and newly obtained information, some conclusions and possible astrophysical
interpretation of the highly unusual astrometric and photometric properties of \st are given in Section \ref{con.sec}.

\section{Pre-whitening based on PCA}
\label{pca.sec}
A number of adverse instrumental effects can cause apparent motion of stellar images across the plane of detector.
They include pointing shifts, gyroscope zero-point crossing, micrometeoroid strikes, etc. The resulting translational (drift)
and rotational (roll) perturbation affects all stars within one CCD channel in very similar ways. Normally,
each target was observed with a fixed digital aperture, or window, comprising a set of specific pixels. The
shape and size of the aperture depend on the point spread function (PSF) and the configuration of near neighbors,
maximizing the collected flux but minimizing blending with other sources.  The Kepler telescope design was
driven by the requirement to observe a large number of targets in a wide field of view almost uninteruptedly.
The Schmidt design with a corrector plate and a mosaic field flattener provides the required field, but the
optical quality of star images is uncharacteristically low for a space telescope.  The wide and often distinctly
asymmetric PSF spills over the edges of the digital aperture, and any motion of the photocentre with respect to
the grid of pixels results in variations in both the aperture flux and the calculated flux-weighted centroid.
Due to the very high sensitivity and relative precision of Kepler measurements, even small amounts of drift
or roll result in correlated changes in the recorded flux and centroids of targets that are situated close
to each other in the detector plane. Differential aberration of starlight\footnote{The main part of stellar aberration 
was taken out by the active pointing control.}  within the wide field of view is
another significant contributor to such correlated trends.

Let ${\bf d}_i$ be a column vector comprising data of interest (for example, a light curve time-sequence) for star $i$.
We can construct a data matrix ${\bf D}$ consisting of vectors ${\bf d}_i$ for a set of stars as long as the observations
are on the same cadence, which is the case with the Kepler long-cadence data\footnote{In fact, the recorded
times of observation may spread by $\sim 0.144$ min between stars on the same chip at the same readout, but this
spread is negligible in the framework of this paper.}. The matrix ${\bf D}=({\bf d}_1,{\bf d}_2,\ldots,{\bf d}_n)$,
where $n$ is the number of objects, has a full rank $n$ and a singular value decomposition (SVD, \citet{gol}):
\eb
{\bf U}\,{\bf \Sigma}\,{\bf V}^{\rm T}={\bf D}.
\ee
The matrices ${\bf U}$ and ${\bf V}$ comprise basis vectors in the space spanned by the columns and rows of ${\bf D}$,
respectively. Only the columns of ${\bf U}=({\bf u}_1,{\bf u}_2,\ldots,{\bf u}_n)$, called the {\it principal
components} of ${\bf D}$, are of interest for this analysis. ${\bf \Sigma}$ is a diagonal matrix comprising
singular values $\sigma_1,\sigma_2,\ldots,\sigma_n$, always in descending order by design, $\sigma_1\ge\sigma_2\ge\ldots
\ge\sigma_n$. Since the norm of the projection $||{\bf D}^{\rm T}{\bf u}_i||={\bf u}_i^{\rm T}{\bf U}\,{\bf \Sigma}\,
{\bf V}^{\rm T}{\bf V}\,{\bf \Sigma}\,{\bf U}^{\rm T}{\bf u}_i=\sigma_i^2$, for any $i$ between 1 and $n$, the first
singular vector ${\bf u}_1$ is the most significant principal component for the given set of data vectors, i.e., the best aligned with
the data vectors, followed by ${\bf u}_2$, ${\bf u}_3$, etc., in decreasing significance. Any of the observations
${\bf d}_i$ can be exactly represented as a linear combination of $n$ singular vectors ${\bf u}_i$, but if we want
to produce the closest and most faithful fit with a limited number of basis vectors, we should take the first ones.
If the data vectors were all orthogonal (which never happens), the singular values would all be equal, and the singular vectors would have the same significance. If there is some degree of commonality in the data vectors, such as common
trends, localized perturbations, etc., the first few singular vectors (or principal components) will represent them
to the highest possible degree. This is essentially the justification of the PCA.

Our objective is to remove as much of the correlated trend in a time series as possible without distorting the genuine
signal specific to a given object. De-correlation of data, which is sometimes called pre-whitening, can be achieved by
removing the most significant principal components from it. The pre-whitened data vector is then
\eb
\hat{\bf d}={\bf d}-\sum_{i=1}^k \alpha_i{\bf u}_i
\label{fit.eq}
\ee
where $\alpha_i$ are the direction cosines of ${\bf d}$ in the subspace spanned by the first $k$ singular basis vectors.
Since the basis vectors ${\bf u}_i$ are orthonormal, these coefficients are computed by direct projection rather than by
a least-squares regression:
\eb
\alpha_i={\bf u}_i^{\rm T}\,{\bf d}.
\ee

The technical implementation of this method is rather subtle, despite the seemingly simple principle.
One of the crucial decisions, which potentially can greatly alter the result, is how to normalize the data
vectors before computing the singular vectors and the principal components. Normalization is required, because
bright stars with a large degree of intrinsic variability can get too much weight in the SVD, which defeats
the purpose of pre-whitening. Our ad hoc method developed for this analysis is similar to the cotrending technique
developed by Tom Barclay, realized as the {\tt kepcotrend} subroutine in the PyKE contributed software package
\citep{sti}. There are important differences in the implementation. The PyKE algorithm is based on principal
components computed for only 50\% of ``most correlated" light curves observed within each CCD channel over a quarter.
We consider it important to use as many data vectors as possible for the SVD, to minimize the influence of intrinsic
photometric variability and reduce the level of noise in the principal components. Therefore, we are using all
light curves collected within a given channel\footnote{It is not advisable to include data from other channels
because the instrumental effects can differ for each particular chip.}. After the selection of most correlated
data vectors, the PyKE algorithm does not apply any normalization for the final SVD, whereas we apply the required subtraction of the median flux followed by a minmax normalization bringing the range of variation to unity. This allows us
to minimize the adverse impact of stars with very noisy data, large degrees of intrinsic variability, or occasional
observational outliers. Finally, we apply the same number of first principal components for all batches of observations.
It is hard to justify skipping some of the most significant components in favor of less significant ones, so the
only choice in this case is how many of the components to use. 

\begin{figure}[htbp]
\epsscale{0.55}
\plotone{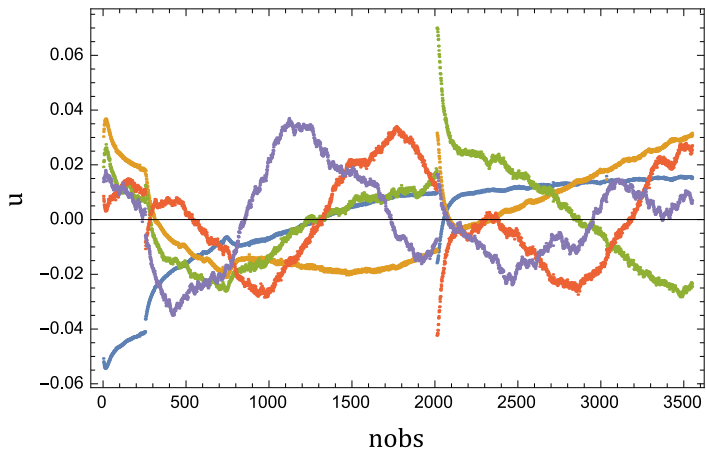}
\hspace{2pc}
\caption{The first five principal components (basis vectors ${\bf u}_i$, $i=1,2,\ldots,5$) of the collection of
1828 light curves collected in Q16 on channel 56. Note that the components capture a common dip in the light
curves at approximately $nobs=2000$, which was probably due to a temporary misalignment of the telescope caused by
a micrometeoroid hit.}
\label{5pc.fig}
\end{figure}
The limiting number $k$ was determined by experimenting with real Kepler light curves. We investigated in depth
the properties of PCA and filtered data for one quarter, Q16, and all 1828 stars observed in this quarter in
channel 56. Our object of investigation, KIC 8462852, was also observed in this channel in Q16. The first 5 principal
components for the collection of 1828 light curves are shown in Fig. \ref{5pc.fig}. Note that the vectors are
orthonormal by design, so they are dissimilar to one another to the highest degree. Generally, they display
some smooth variations with time, but also capture some short-term adverse effects. After applying
progressively greater numbers of principal components according to Eq. \ref{fit.eq},
and computing the 2-norm of residuals, we determined that the latter begins to be reduced by 1\% or less
with respect to the most significant component
already when 4 principal components applied. Consequently, the filtered light curves with 5, 6, etc. components
applied were hard to distinguish from the 4 PC-filtered curve by eye. The amount of high-frequency noise, which is
present in the principal components in small amounts, on the other hand, grows roughly proportional to the number
of PC used. Therefore, we found it nearly optimal to use 4 components throughout our analysis.

We also checked if a few stars with large degrees of variability can affect the first 4 principal components,
a potentially harmful possibility which can bring about unintended consequences. Our scaling and normalization
scheme turns out to be quite efficient in downweighting such extreme outliers. The first principal components do not
display any features from the most variable data vectors, which are usually pulsating, eclipsing stars or artifacts.
The same is true for the principal components of the $x$ (column) and $y$ (row) coordinates. For example, by far
the largest variation in column pixel coordinates for the test sample, exceeding 2 pixels in magnitude, is found for the target KIC 8653134. These great excursions of the photocenter are artifacts caused by blending with some extraneous
very bright signal, also observed in Q4 and Q8. Without the normalization of observed data by the standard deviation
of the observed variation, this single data vector could have corrupted the computed principal components and damaged
the PCA pre-whitening.
\begin{figure}[htbp]
\epsscale{0.55}
\plotone{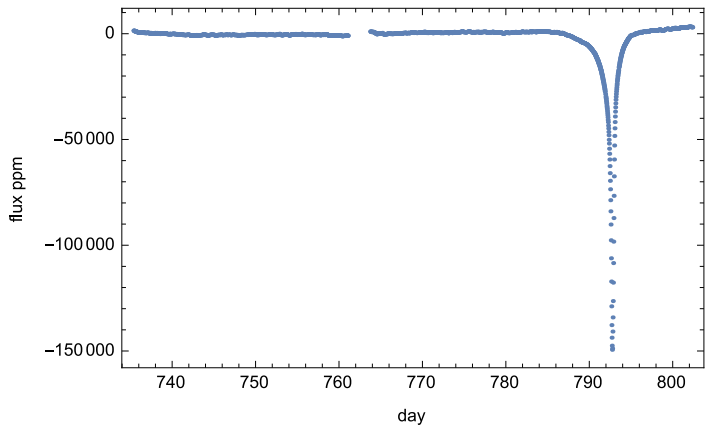}
\hspace{2pc}
\caption{PCA-cleaned light curve of \st in Quarter 8 in relative units of ppm. 
The major flux dip on day 793 is one of the most prominent and unusual features of this dataset almost reaching 15\%.}
\label{q8lig.fig}
\end{figure}
\begin{figure}[htbp]
\epsscale{1.1}
\plottwo{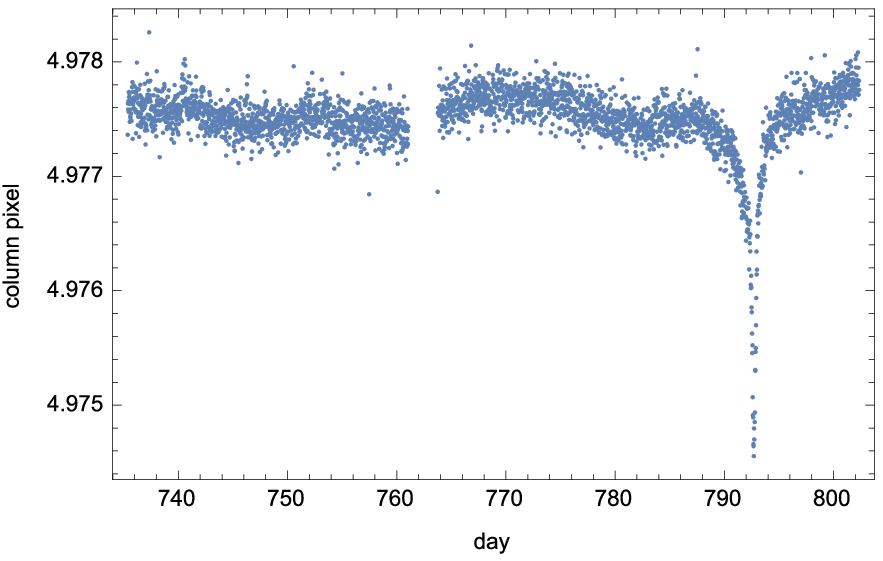}{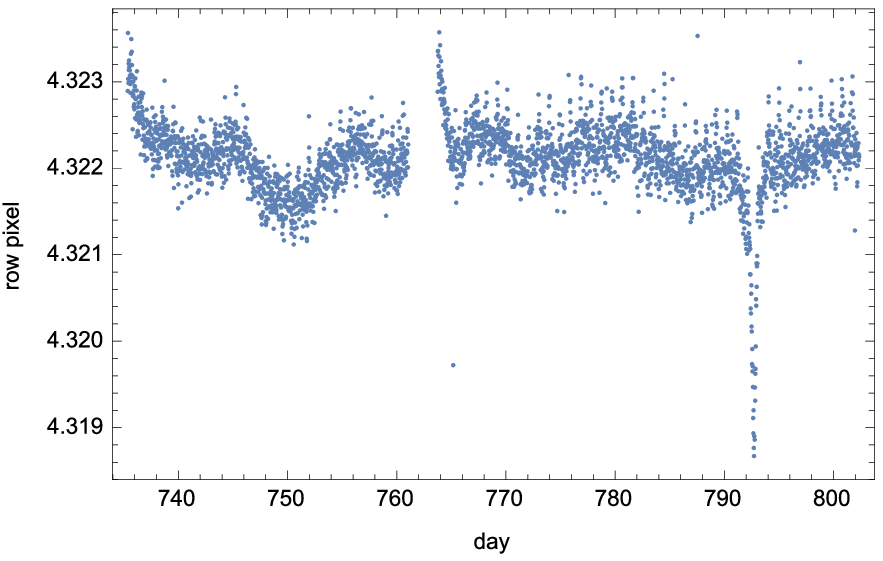}
\hspace{2pc}
\caption{PCA-cleaned centroid trajectories of \st in Quarter 8 in column (left) and row (right) pixel coordinates. 
Note the clear VIM effect caused by the major flux dip on day 793 and the periodic motion in row pixels with 0.88 d period.}
\label{q8pix.fig}
\end{figure}

Another important difference with the PyKE implementation of the PCA is the way we perform the fitting of
the principal components to the data. Eq. \ref{fit.eq} shows a simple projection of the data vector onto the
null space of the first basis vectors. For many regular stars of moderate or small variability, this simple
method works fine, but targets with powerful changes in the light curve, such as our target KIC 8462852, may require
a more careful selection of the data points to fit. A single dip, such as the one on day 793, can push the fit
giving too much weight to a basis vector that has a depression in this part of the time series. The PyKE
PCA algorithm employs iterative filtering of data points deviating by more than a certain threshold amount
from the fit in the preceding iteration (clipping of outliers). We chose a more advanced method of dealing
with outliers, which we had successfully used in other analyses. Generally, a least-squares (LS) method heavily relies 
on the assumptions that are often not met in practice. In particular, it is assumed that data errors are normally 
distributed. If there are outliers in the data (as usually is the case), a single  outlier at a few standard deviations from the mean can strongly perturb the fit. This happens because of the sensitivity of the estimator (2-norm
of the residuals vector) to extreme outliers which may reflect a true signal as well as statistical flukes.
In the latter case, using a 1-norm minimization has been proposed, but it underperforms when occasional outliers
betray a short-term, powerful physical signal.

The previously mentioned simple clipping  approach is to perform an initial LS fit, find outliers  at $t \sigma$ from 
the mean ($\sigma$ is standard deviation of post-fit normalized residuals $r$, $t$ is a threshold, usually equal to 3) and repeat the 
fit without those. This is equivalent to minimizing the sum of functions $\psi(r)$, where  $\psi(r)=r^2$ for $r<3\sigma$
 and 0 otherwise. Multiple other robust functions were suggested in the statistical literature \citep{hub}, Tukey 
bisquare  being
one of the most popular. For small residuals it is identical to the square of residuals (just like in LS method), but it tapers off for 
more statistically significant residuals according to a smooth weight
function. The Tukey bisquare
regression estimator reaches the 95\% efficiency for normally distributed data at $k= 4.685$, and a breakdown point of 0.5 of the S-estimator is reached at $k=1.548$. 
Numerous popular packages such as R, SciPy for Python, or Matlab, provide prepackaged  procedures  for fitting a robust regressor with various estimator functions, including Huber, Tukey, and trimmed LS.

\section{Control star KIC 3836439 = HIP 93954 = HD 178661}
\begin{figure}[htbp]
\epsscale{1.1}
\plottwo{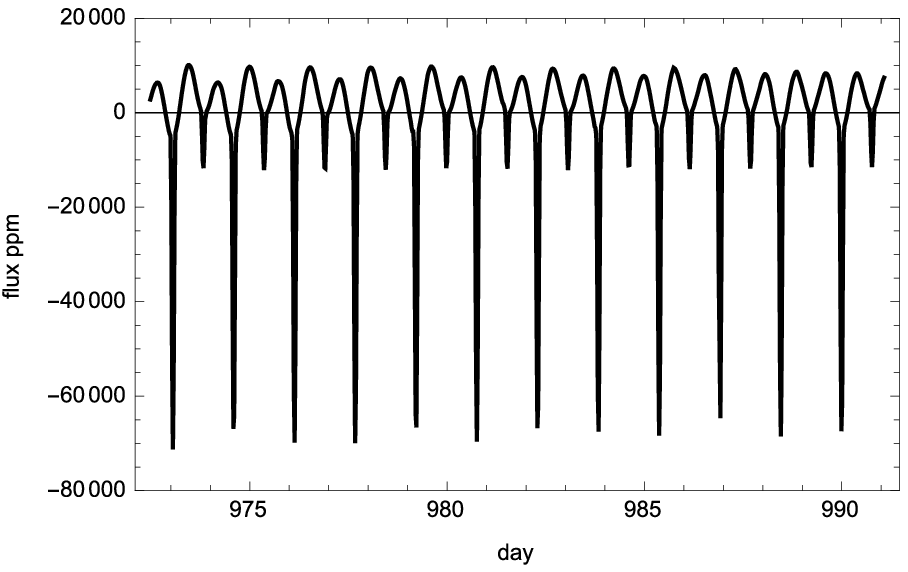}{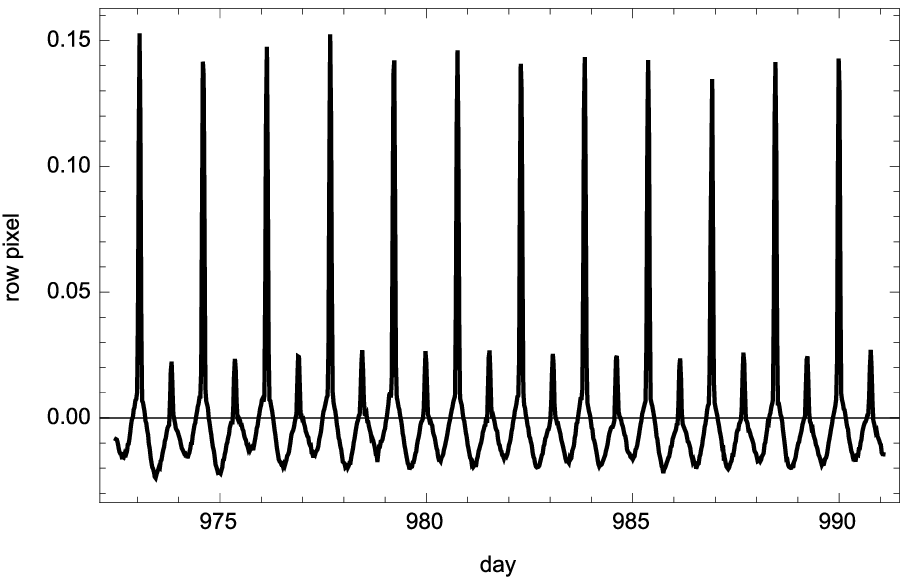}
\hspace{2pc}
\caption{PCA-cleaned normalized SAP flux (left) and centroid trajectory of our control star KIC 3836439 for a segment of the main
mission data.}
\label{con_fx.fig}
\end{figure}

To validate our PCA pre-whitening technique and to test the sensitivity of Kepler astrometry and photometry, we randomly selected the control
object HIP 93954 from the list of known eclipsing variables \citep{sla}. It is also a powerful VIM object confidently detected in all
the 17 quarters of the main mission \citep{mak}. Fig. \ref{con_fx.fig} depicts a segment of the SAP light curve (left) and the relative
centroid position in row coordinates (right), both data series pre-whitened and cleaned as described in Section \ref{pca.sec}. The VIM
effect is evident from the close similarity of the photometric and astrometric variation.

First, we note that the pre-whitening procedure has not affected the intrinsic variability pattern of this eclipsing star in
either astrometry or photometry but removed the long-term perturbations. The centroid trajectory looks like a mirror reflection
of the light curve. A periodogram analysis performed on both curves confirmed that the main signal resides at a period of
$0.77$ d, which is half the orbital period as determined by Kepler ($P_{\rm orb}=1.5404$ d) and by previous spectroscopic studies.
HIP 93954 is a multiple system, at least a triple, with a B companion separated by $1\farcs31$ from the primary at 1991.25,
as determined by Hipparcos \citep{esa}. The separate Tycho-2 photometry for the A and B components is, respectively,
$V_T=7.96$, $B_T-V_T=0.20$, and $V_T=9.22$, $B_T-V_T=0.47$ mag \citep{fab}. The flux from the primary and the secondary, residing within
the same pixel on the Kepler CCD, is totally blended. Both the astrometric and photometric data contain a fair amount of signal
from the B component, which is practically unknown.

\begin{figure}[htbp]
\epsscale{0.55}
\plotone{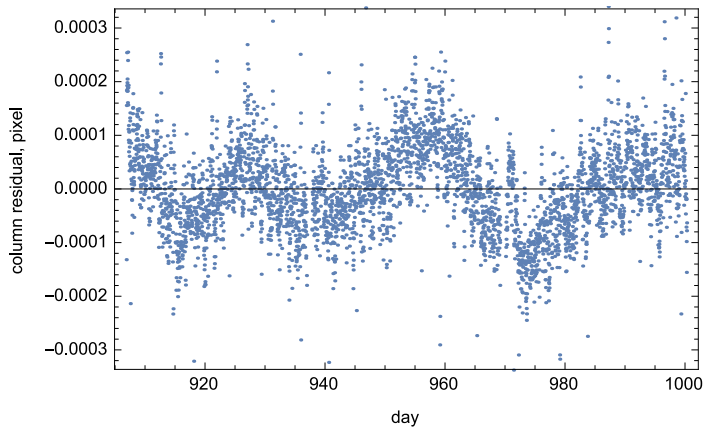}
\hspace{2pc}
\caption{Residual centroid trajectory of the control star KIC 3836439 after fitting out the correlated VIM effects (see text).}
\label{res_x.fig}
\end{figure}

This star is also listed in the UrHip catalog \citep{fro} with a proper motion which is statistically different from the short-term
proper motion determined by Hipparcos. This likely indicates a measurable astrometric orbital motion at the level of a few mas yr$^{-1}$.
Are our PCA-cleaned centroid trajectories good enough to detect the orbital motion directly? To separate the astrometric signals
related to VIM from other, possibly intrinsic, astrometric changes, we performed the following modeling of data. If $\bf{f}$ is
the PCA-cleaned light curve (flux), $\bf{x}$ is the centroid coordinate on the same time cadence $\bf{t}$, the linear model for
$\bf{x}$ is obtained by solving the Least-Squares problem
\eb
[{\bf 1}\: ({\bf f}-f_0)/{\bf f}]\; {\bf a}={\bf x},
\label{model.eq}
\ee
where $f_0$ is the mean flux outside of the photometric event (quiescent total flux).
A fit $\hat{\bf x}$ can be obtained for a quarter ($\sim 90$ days) or a smaller segment of data.
We are interested in the residuals ${\bf x}-\hat{\bf x}$ which should be clean of the powerful correlated VIM signals. Fig. \ref{res_x.fig}
shows modeled residuals for the column coordinates for an entire quarter of the mission. Here we achieved a significant reduction
of astrometric variations to small fractions of one milli-pixel. The residual trajectory is very flat and has a sample standard deviation
of 0.34 mas in sky angle units. The remaining pattern, including the distinct variation with a period of approximately
0.5 d, is likely another VIM signal coming from the
blended sources. 

This example shows the power and potential of astrometric analysis and VIM processing of Kepler data. The light curve variations are
dominated by the brighter blended companion, which is normally the target star. The variability of the fainter companion is also
present in the light curve but it is swamped by the flux of the primary. In most cases, it is impossible to separate signals coming from
the blended sources based on just the flux variations. Because of the nature of VIM (Section \ref{vim.sec}), the astrometric part of the data
includes much stronger and readily discernible signal coming from the faint companion. As in this case, it becomes evident especially after
removing the main correlated part from the astrometric trajectory. This technique allows us to separate blended variability signals and
investigate the properties of faint unresolved companions in some cases.

\section{Photometric and astrometric events}
\label{eve.sec}
Returning to the object of our investigation \st, a few major dips
detected in the 17 quarterly long-cadence light curves are listed in Table \ref{eve.tab}. Two events, the first on day
793 and the second on day 1519, stand out due to their surprising amplitudes of $\sim 20$ and $\sim15$\%, respectively. The amplitudes of other
dips range between $\sim1000$ and $\sim70000$ ppm. The timing of these events does not seem to be commensurate with any periodic
occultation, invalidating the eclipsing binary hypothesis. Some of the dips, but not all, seems to be preceded and followed by
smaller amplitude brightening events, but the physical reality of them is hard to ascertain from the filtered data. \citet{mon}
find no evidence for most of them. The first order of business is to determine if these signals come from the intended target 
or from a blended source. 

Images with bright sources (e.g., Jupiter) obtained with Kepler show extended areas of light contamination spilling over far beyond 
the typical digital aperture. These extended PSF structures are caused by the complicated patterns of stray light (which is time-variable), 
internal reflection and diffraction, plus the unfortunate shutter-less mode of readout. The large-scale PSF is non-axial symmetric. 
For these reasons, the contribution of light from distant sources (well outside of the digital aperture) is bound to be time-dependent and orientation-dependent. With the data in hand, it is impossible to decisively separate situational, long-range contamination
signals from more permanent, close range blended confusion sources, especially when the photometric excursions are relatively
sparse as in our case. An exposition of possible causes for long-range contamination of photometric data was given by \citet{cou}.
Some of the contamination modes, such as the column anomaly, can be ruled out in our case, but others (e.g., a ghost image due to internal
reflection) may still be possible for the weak signals in the light curve. 
As discussed in the following, the VIM response appears to be nonlinear (a relatively smaller VIM speed is
measured for larger photometric signals, and vice versa), which may be a hint at a distant blended source. But a tight companion can produce quarter-dependent VIM speeds too due to the non-axial, irregular shape of the PSF. 
 \begin{table*}
 \centering
 \caption{Photometric dips in the light curve of \st.}
 \label{eve.tab}
 \begin{tabular}{@{}lrcl@{}}
 \hline
            &                 &       &\\
   Day     &  Amplitude    & VIM? & Notes\\
           &  ppm          &      &\\
 \hline
 140 & 4000 & yes  &  brightening before and after event by +1000 ppm\\
 215 & 1000 &      & modulation in row pixels \\
 260 & 6000 & yes & possible brightening after eclipse \\
 377 & 1800 & ? &  \\
 427 & 2100 &   & \\
 502 & 1200 &   & narrow dip, no obvious brightening before or after\\
 610 & 2400 & yes & not an eclipse but a sudden drop, instrumental effect?\\
 660 & 1000 & & broad, shallow dip, complex variability in this quarter \\
 793 & 150000 & yes & no obvious brightening before or after\\
 1144 & 1000 &  & jitters around probably of instrumental nature\\
 1206 & 4000 & yes & narrow, well defined event\\
 1519 & 204000 & yes & most prominent event\\
 1540 & 25200 & yes & bracketed by brightening events, VIM speed is clearly different\\
 1568 & 74500 & yes & bracketed by brightening events\\
 \hline
 \label{table}
 \end{tabular}
 \end{table*}

\subsection{Blended companions}
\label{second.sec}
\begin{figure}[htbp]
\epsscale{0.95}
\plotone{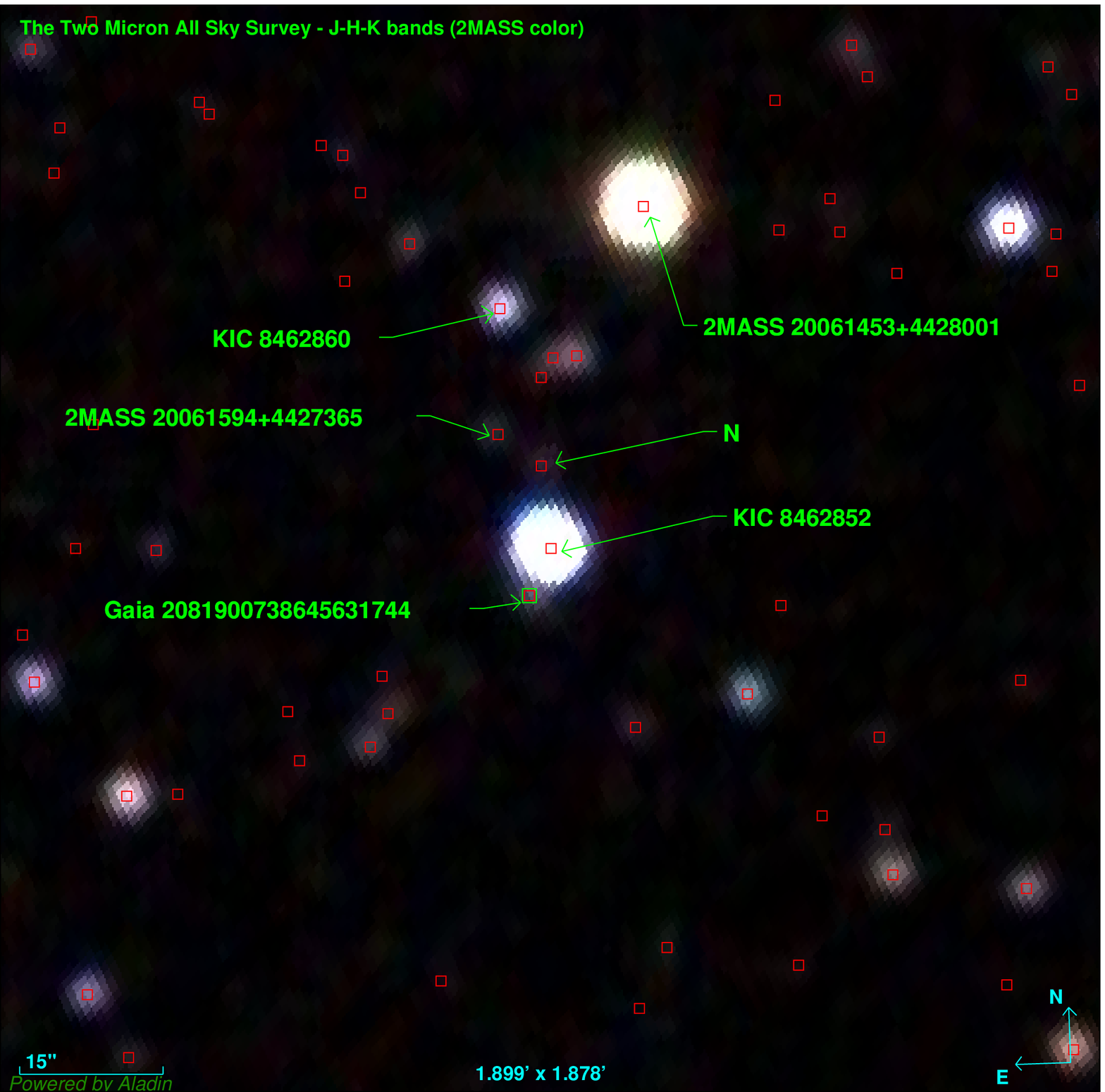}
\hspace{2pc}
\caption{A finder chart of the small area centered on \st with tagged other stars discussed in the text,
rendered from the 2MASS images using the Aladin application from CDS \citep{bon}. All sources listed in Gaia DR1 \citep{bra} are
marked with small red squares. The E companion discovered by \citet{boy} using adaptive optics imaging is not
visible in this image, perhaps being too close to the target star and unresolved.}
\label{finder.fig}
\end{figure}

The closest known companion was resolved by Keck AO imaging at 2 arcsec due east of the target. It is fainter by
3.8 mag in the NIR $H$ band \citep{boy}. Because of the small angular separation and large magnitude difference, this star (called 
E-companion) is not
listed in any major catalog. Regular large-scale images reveal several more distant, but also faint, companions. Fig. \ref{finder.fig}
shows a stacked 2MASS color finder chart of the immediate surroundings of the target star. The most 
significant neighbors are a star approximately north of the target separated by 9 arcsec from it (we will call it N-companion) and fainter by 5.4 mag in $J$, and another star 6 arcsec from the N-companion with a magnitude difference
of $\Delta J= 5.6$ mag. Both these faint neighbors are listed in the major catalogs, including the 2MASS \citep{cut}
as 20061551+4427330 and 20061594+4427365, respectively. There is another somewhat fainter companion approximately
at position angle $150\degr$, separation 6 arcsec from the target, not named or listed in any of the major astrometric
catalogs except in Gaia DR1 as 2081900738645631744. All these chance companions, as well as a few even fainter neighbors, are at least partially blended
with the target in the digital aperture of \st. 

The largest degree of flux blending occurs for the closest known neighbor E. Up to 3\% of the recorded flux may be
coming from this neighbor, and up to 0.6\% of light is coming from the N-companion and other neighbors. Obviously,
the largest dips of 15\% and 20\% of the total flux could not originate from any of the neighbors even in the event
of their complete occultation. The more numerous smaller dips and other photometric features in the observed light
curve could in principle be coming from the blended neighbors, a possibility that should not be outright dismissed.

\subsection{VIMs}
\label{vim.sec}
Comparing the PCA-cleaned light curve of \st in Q8 (Fig. \ref{q8lig.fig}) with the PCA-cleaned centroid trajectories
for the same quarter (Fig. \ref{q8pix.fig}) we immediately see that the target displays VIM effects. The 15\% drop in flux
is mirrored by similarly shaped dips in both coordinates on the detector CCD, $x$ (column) and $y$ (row). In this case,
we know that the photometric event happened on the target star, \st. The moment-based photocenter shifted because of
the dimmed light from the brightest component. As explained in \citep{mak}, the VIM data can be used to estimate
the position angle of the main blended component, but not the separation. The relevant approximate formula for the
separation $s$ is
\eb
s\approx -\left[\frac{\Delta s}{\Delta f}\,f\right]\frac{f+\Delta f}{f_2},
\label{vim.eq}
\ee
where $f$ is the total observed flux outside the photometric event (e.g., the median flux in the ``quiescent" state),
$f_2$ is the flux of the blended ``constant" component, and $\Delta s/\Delta f$ is approximately equal (for small flux variation) to
the measurable parameter $ds/df$ called VIM speed,
which is the ratio of the astrometric excursion magnitude to that of the corresponding photometric excursion.
For the Q8 event, the VIM speed is very small, because a 15\% photometric dip ($df/f\approx 0.15$) caused an astrometric
perturbation of just a few thousandths of a pixel on the CCD. It is owing to the outstanding single measurement
precision of Kepler astrometry (with a standard deviation of $\sim 0.8$ mas on the sky) that such tiny signals can
be confidently detected at all. The separation, however, remains ambiguous because we do not know the fraction
of light from the unperturbed companion. But useful estimates can be obtained for the known companions with the
available order-of-magnitude knowledge of the flux ratio $f/f_2$.

\begin{figure*}
\epsscale{1.1}
\plottwo{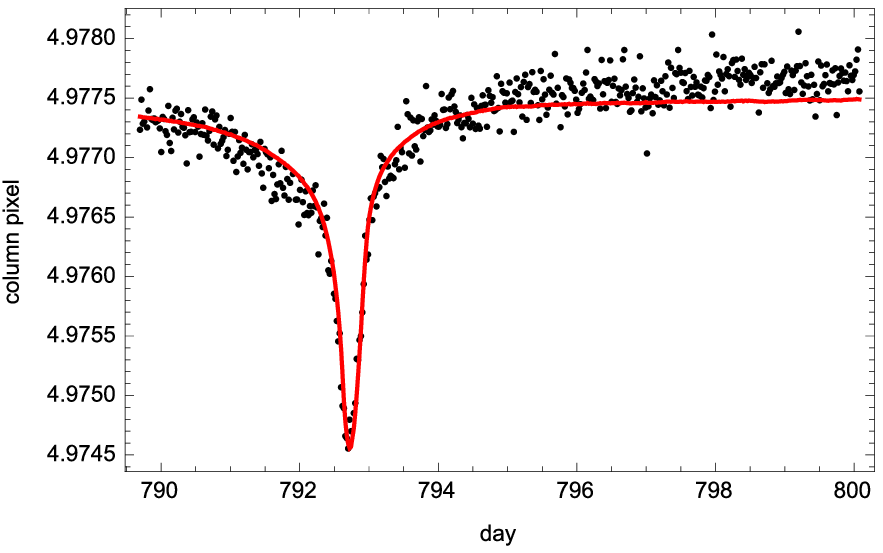}{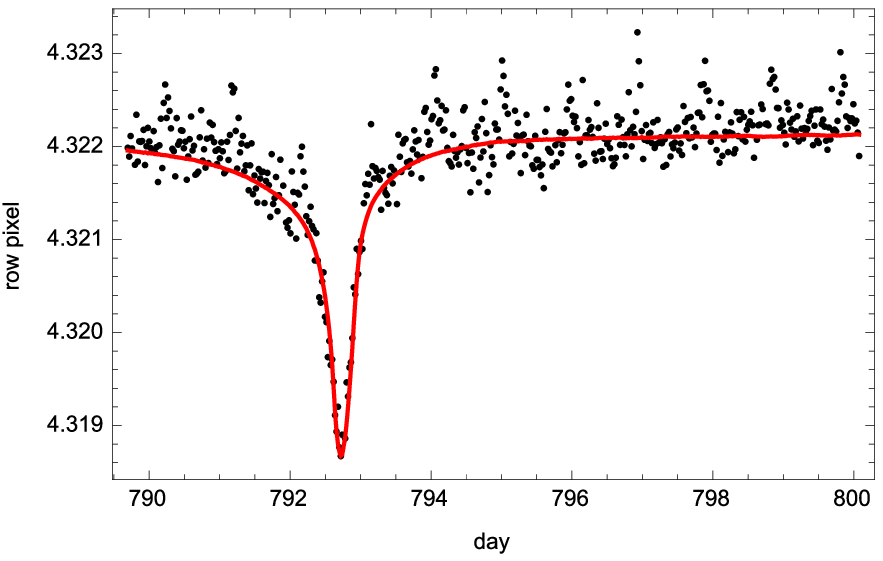}
\hspace{2pc}
\caption{PCA-cleaned centroid trajectories of \st in Quarter 8 in column (left) and row (right) pixel coordinates
around the major photometric event on day 1519, marked with dots. The superimposed solid curves represent the normalized
and rescaled flux observed on the same cadence.}
\label{q8pixf.fig}
\end{figure*}

Fig. \ref{q8pixf.fig} gives a zoomed-in view of the astrometric trajectories around the photometric dip on day 793 (dots) superimposed with the observed light curve (solid line), which was rescaled and normalized to achieve the best match.
The degree of conformity between the astrometric and photometric signals during the 15\% dip in flux is remarkable.
The normalization coefficient applied to the light curve yields the VIM speed, estimated at $-7.9\times 10^{-8}$
pix$/$(e s$^{-1}$) for columns and $-9.2\times 10^{-8}$
pix$/$(e s$^{-1}$) for rows of the CCD 56. The VIM speed magnitude is thus $1.2\times 10^{-7}$
pix$/$(e s$^{-1}$) in the direction $229\degr$ on the CCD. With a median flux of $2.4\times10^{5}$ e s$^{-1}$,
the coefficient in the square brackets in Eq. \ref{vim.eq} amounts to $2.9\times10^{-2}$ pixels. The closest companion
E is at 2$\arcsec = 0.5$ pixels. Hence, the estimated ratio $f/f_2$ is roughly 17, if companion E is responsible
for the VIM. This number seems to be small given the magnitude difference $\Delta H=3.8$ mag, unless companion
E is a blue star. On the other hand, the closest neighbor may not be the dominating VIM companion, because the
VIM speed, which defines the magnitude of the astrometric signal, is proportional to the separation. Somewhat
counter-intuitively, a more distant neighbor may cause a greater VIM as long as a significant fraction of its
flux falls within the aperture. Furthermore, a distant bright star situated outside the aperture may still be
responsible for the largest VIM even when only a small fraction of its flux falls within the aperture, because of
a different distribution of light in the extended wings of the image (in which case Eq. \ref{vim.eq} is not applicable).

Using the WCS constants and the pixel scale parameters specified in the metadata of the long-cadence FITS files,
we transform the VIM direction ($229\degr$) in the column-row pixel coordinates into the conventional position angle
on the sky, reckoned counterclockwise from north through east. This transformation yields a PA$=336\degr$ with
an uncertainty of about $\pm 10\degr$. This direction does not match any of the previously discussed close companions.
However, an interesting star is located at a larger distance from the target, viz., $36\arcsec$, PA$=344\degr$ (Fig.
\ref{finder.fig}). It is
listed in several major catalogs, including 2MASS as 20061453+4428001. The star with a $V\approx 15.5$ mag is optically
fainter than its neighbor \st by 3.6 mag, but the 2MASS magnitudes $J=10.577$, $H=9.455$, $K_s=9.075$ indicate that it is
exceptionally red and in fact brighter in the infrared. Little is known about this star apart from the NIR and MIR magnitudes and proper motion (which is small and unremarkable). The J2000 ICRS coordinates of the star are RA
$=301.560574\degr$, Dec$=+44.466687\degr$ in the URAT1 catalog \citep{zac}. 

One may wonder why this distant and faint companion produces the weak, but measurable, VIM effects while the closer neighbor
KIC 8462860, discussed by \citet{mon} as a likely contamination source, does not? KIC 8462860 is also 3.65 mag fainter than our
target star, separated by 25 arcsec at PA=$\simeq 20\degr$. One possible explanation would be that the bloated and strongly
asymmetric PSF in this corner of the Kepler field of view, depicted in Fig. 1 of \citep{mon}, picks up more photons from 
the sources in specific directions on the CCD. However, we will see that the VIM direction of the major events is consistent
between the mission quarters, where the orientation of the CCD and the PSF was different on the sky. This interpretation meets 
significant difficulties, leaving room for a very close, yet unresolved, companion.
\begin{figure*}
\epsscale{1.1}
\plottwo{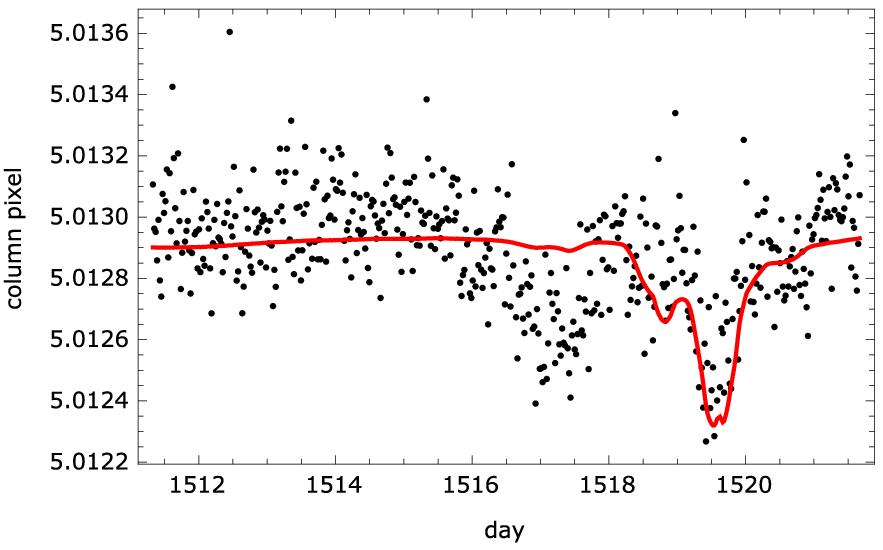}{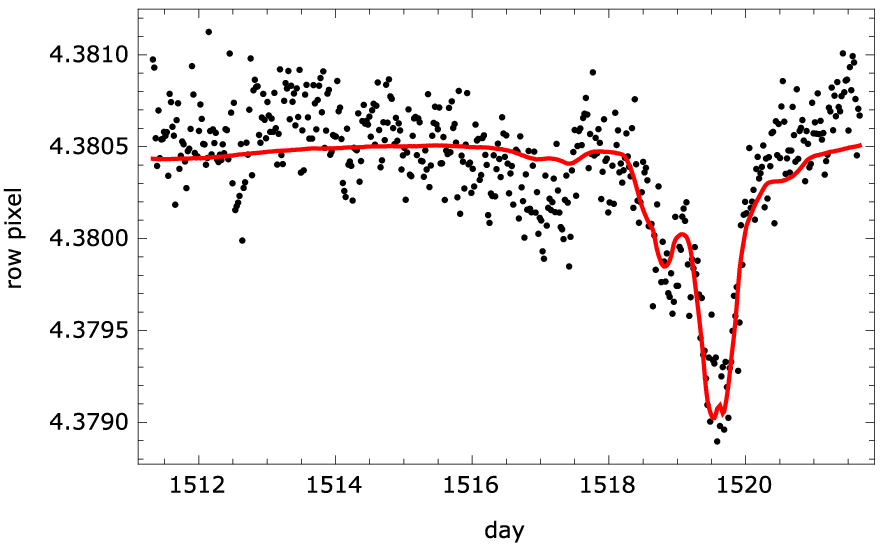}
\hspace{2pc}
\caption{PCA-cleaned centroid trajectories of \st in Quarter 16 in column (left) and row (right) pixel coordinates
around the major photometric event on day 1519, marked with dots. The superimposed solid curves represent the normalized
and rescaled flux observed on the same cadence.}
\label{q16pixf.fig}
\end{figure*}

Speculatively, if the weak VIM signals are caused by a very close unresolved companion, it should be very close to the target star,
and also rather faint. Given the limits of the high-resolution imaging and speckle interferometry from \citep{boy}, it should be
either closer than $\sim40$ mas, or have a delta magnitude above 4. Furthermore, it should be an optical companion given the
absence of radial velocity variations, i.e., a chance projection on the sky of an unrelated star. The probability of such projections
is quite low, of order two parts in $10\,000$. But between the $170\,000$ Kepler targets, there should be a few dozens such alignments.
Some of these alignments are bound to include foreground companions. This may give some more credibility to the hypothesis of
chance occultations by foreground objects presented in Section \ref{junk.sec}. 

The major event on day 1519 took place in Q16, almost exactly two years after the 15\% dip on day 793. The Kepler
telescope observed the same area of the sky in four different and fixed orientation states (roll angle). Each
orientation state was strictly repeated in a one-year cycle. Therefore, the target was observed in Q16 with
the same orientation of the telescope, and in fact, on the same pixels, as in Q8. We therefore expect to
find a VIM effect similar in direction and magnitude to the one on day 793. The observational data thwart this
expectation and provide additional puzzles.

Fig. \ref{q16pixf.fig} displays a zoomed-in portion of the astrometric data (black dots, left plot for $x$
and right for $y$) superimposed with the appropriately shifted and scaled light curve (solid line). The main dip is split,
with a smaller component preceding the larger one by some 18 hours. Both components are reproduced in the centroid
trajectory, but apparently not as closely as in Fig. \ref{q8pixf.fig}. The estimated VIM speed is 
$-1.3\times 10^{-8}$ pix$/$(e s$^{-1}$) for columns and $-3.1\times 10^{-8}$ pix$/$(e s$^{-1}$) for rows,
with a magnitude of $3.4\times 10^{-8}$ pix$/$(e s$^{-1}$) and a position angle of $247\degr$ on the detector.
Even though the direction is roughly the same as in Q8, the magnitude of VIM is smaller by a factor of 3.5.
The observed nonlinear character of these VIM effects may be explained by the influence of
a very distant, but bright neighbor. The brightest companion at large is the rather unremarkable red giant star
KIC 8462934 ($V=11.51$ mag) separated by 89 arcsec, 
which was investigated for asteroseismology \citep{pin} but does not exhibit any unusual
variability. The detected VIM speed in Q16 can be significantly different with the same configuration of blended images,
if the light distribution of the portion coming from the distant neighbor has a significant gradient, and the position
of the aperture with respect to the pixel grid is slightly shifted with respect to that in Q8. Alternatively, we may
conclude that the astrometric excursion is capped at approximately 12 mas on the sky, no matter how deep the underlaying photometric
variation is. This explanation would be consistent with the unresolved optical companion hypothesis.

Fig. \ref{q16pixf.fig} provides an even greater puzzle outside of the main event 1519. We can clearly see a dip
in both coordinates on day 1517, which barely registers in the re-scaled light curve. The excursion is larger
in the column direction, where it is comparable in magnitude to the main VIM effect, but the observed total flux
hardly changed at that time. This implies that the VIM speed of this event is more than 10 times greater than the
VIM speed of the main photometric event, and the direction is somewhat different too. How can the VIM speed be
so different for two close VIM occurrences with a fixed configuration of stars? Eq. \ref{vim.eq} implies that the VIM
speed is proportional to $f_2/f$, where $f_2$ is the flux from the photometrically constant companion. If a much
fainter companion is dimmed while the bright target star remains constant, the ratio $f_2/f$ is close to
unity, and the VIM speed can be greater by orders of magnitude even for the same pair of blended images. In other
words, the VIM speed is extremely sensitive to whether the photometric change happens on the bright or the faint 
companion. The conclusion is that the VIM on day 1517 was caused by a photometric variability of {\it another} star.

Given these observations, we begin to suspect that the seemingly disorderly collection of photometric dips is coming
from different sources overlapping with the aperture footprint. 

\subsection{The origin of small-amplitude dips}
\label{small.sec}
\begin{figure}[htbp]
\epsscale{0.55}
\plotone{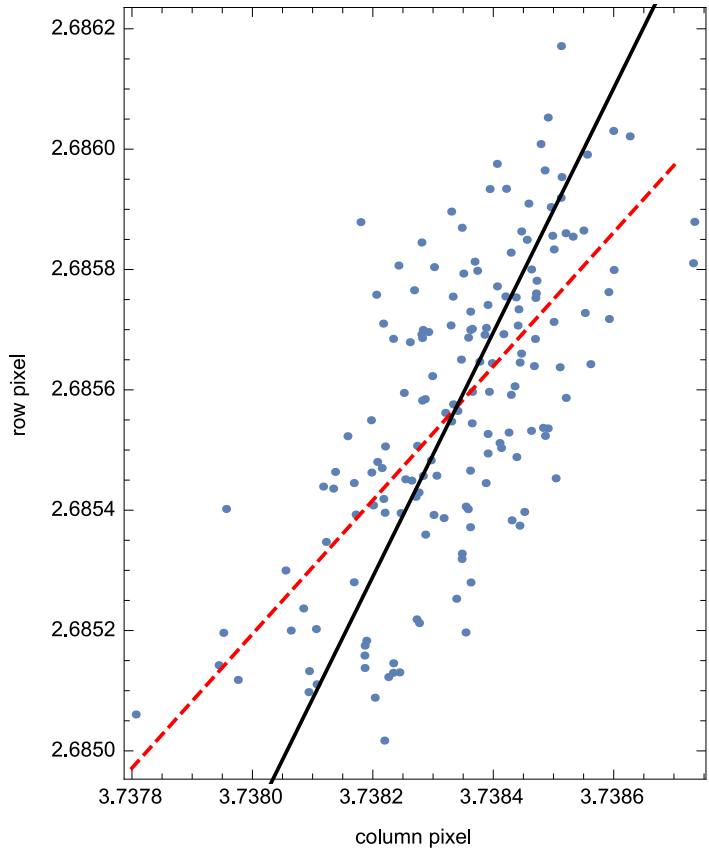}
\hspace{2pc}
\caption{The segment of the VIM trajectory in detector's pixel coordinates caused by the small-amplitude
photometric event on day 260 in Q3. The regular least-squares fit is shown with a dashed line; the more
accurate orthogonal distance regression is shown with a solid line.}
\label{xy.fig}
\end{figure}

The smaller dips in the light curve at a few to several thousands ppm observed across the mission time span
are sometimes associated with marginally small excursions in centroid position (Table 1). These barely detectable
VIM effects require a more sophisticated method of analysis than the previously used rescaling and amplitude
matching. The astrometric trajectories are especially difficult to analyze for small-amplitude wobbles because
of the presence of larger perturbations on similar or longer timescales. We focus in this paper on two medium-sized
events on days 140 (Q1) and 260 (Q3), which are the most distant in time from the two major events at the end of the
mission. 

For each of these VIM events, we select time boundaries corresponding to the beginning and the end of the flux dip.
For example, for the first event, the interval is $[138.3,143.6]$ days. We use only the PCA-cleaned data within
these intervals. If we plot coordinate measurements versus simultaneous flux measurements, a correlated VIM shows up
as a linear dependence with a slope indicating the VIM speed $ds/sf$. If we plot $y$-coordinates versus $x$-coordinates
for the same cadence of measurements, the slope of a linear trend indicates the direction of the correlated 2D
motion, i.e., the VIM direction. Fig. \ref{xy.fig} shows the $y$ versus $x$ plot for the stronger and more discernible
VIM on day 260. The next step is to fit a straight line to this cloud of points. The regular least-squares linear
regression, shown with the dashed line in the plot, provides a poor fit in this case strongly underestimating the
slope. This is because the least-squares (LS) method is designed to minimize the quadratic sum of residuals in the
fitted variable ($y$), while the argument values ($x$) are implicitly assumed to be precise. In our case, both
coordinate measurements are perturbed by random errors. The more appropriate method of fitting is the orthogonal
distance regression, described in many statistical textbooks. The simplest algorithm in implementation is the total
LS, which allows to fit 2D data sets of inhomogeneous precision and to employ elements of robust
statistical estimation (including Tukey bisquare estimator) via appropriate weighting of data points, but it performs
only the 2-norm optimization. The solid line in Fig. \ref{xy.fig} shows thus obtained total least-squares fit assuming
the same error in $x$ and $y$. Visually, it provides a much more realistic fit given uncertainties in both coordinates.
We also tried a generalization of the orthogonal distance regression adapted to the 1-norm solution (i.e., minimizing
the absolute values of weighted residuals in both coordinates), but found the results close to the total LS fit. 

Using the appropriate WCS constants given in the metadata, we estimate the VIM directions at $336\degr$ for Q1
and $329\degr$ for Q3. These position angles are consistent, within the expected errors, with the estimated directions
of the two major events (Section \ref{vim.sec}). The most plausible interpretation is that the same faint companion,
possibly 2MASS 20061453+4428001, is mostly responsible for the observed VIM and the dimming occurs on the target
star \st. Indeed, if the dimming belongs to the companion, the photocenter should move in the opposite direction.
The VIM speed magnitude, on the other hand, is less consistent with this scenario. The estimated values are
$3.6\times 10^{-7}$ pix$/$(e s$^{-1}$) in Q1 and $1.4\times 10^{-7}$ pix$/$(e s$^{-1}$) in Q3. The observed scatter in
VIM speed between individual events is hard to explain. In particular, the estimated VIM speed for the major
event on day 1519 is roughly 10 times smaller than that for the much smaller dip on day 260. It appears that
no matter how large the photometric changes are, the astrometric excursion is upper-limited to $\sim 12$ mas on
the sky. If this is true, the only possible explanation is that there is an unresolved companion at a comparably
small separation.

\subsection{Medium-amplitude photometric events}
\begin{figure}[htbp]
\epsscale{1.1}
\plottwo{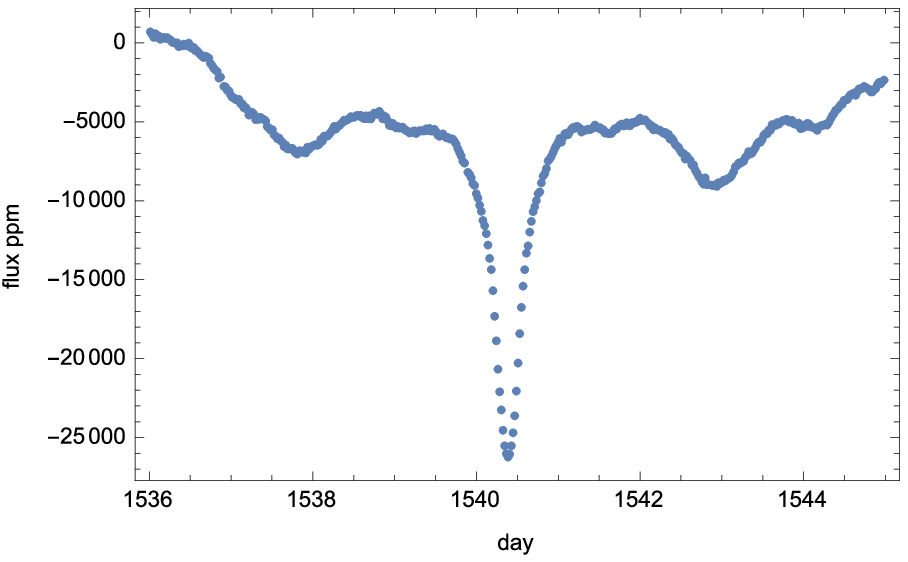}{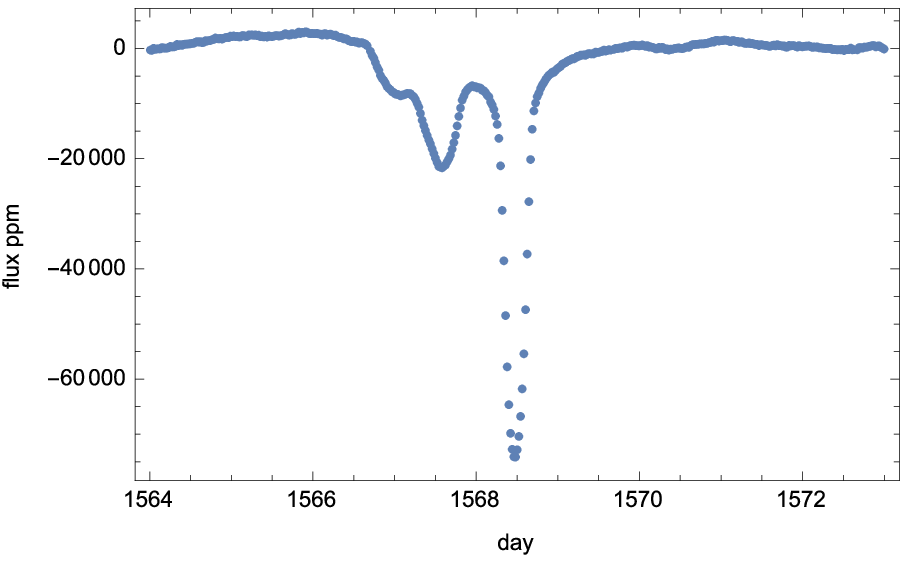}
\hspace{2pc}
\caption{PCA-cleaned light curves around the medium-amplitude
photometric events on day 1540 in Q16 (left) and day 1568 in Q17 (right). }
\label{q16dip.fig}
\end{figure}

To complete our VIM analysis, we analyze two medium-amplitude photometric dips that occurred on day 1540 (Q16), or
some 20 days after the greatest event, and day 1568 (Q17). Fig. \ref{q16dip.fig} shows the portions of the PCA-cleaned 
light curve centered on these two events. Unlike the two major dips in Q8 (Fig. \ref{q8pixf.fig}) and Q16
(Fig. \ref{q16pixf.fig}), with fairly smooth ingress and egress, both variations in question are marked by pronounced 
brightening preceding and following the dip. We applied the algorithm described in Section \ref{small.sec} to both
medium-amplitude events to determine the VIM parameters

The day 1540 dip in Q16 is studied by using the astrometric and photometric measurements between days 1539 and 1541.
An orthogonal distance regression yields a $y(x)$ slope of 1.564. The corresponding position angle on the sky
is $328\degr$. From $x(f)$ and $y(f)$ fits, we derive a VIM speed of $2.6\times 10^{-7}$ pix$/$(e s$^{-1}$), which is in
the range of small-amplitude events, but significantly greater than the VIM speed of the major events.
The day 1568 event took place roughly between days 1567 and 1570. The position angle is estimated at $322\degr$, and the VIM
speed is approximately $1.2\times 10^{-7}$ pix$/$(e s$^{-1}$).

The six photometric dips we analyzed in detail for VIM effects vary in amplitude from 4000 to 204000 ppm, but they all
have consistent direction on the sky, PA$=330\degr\pm 6\degr$. This points at one of the several faint companions of the
target star as the dominant source of astrometric perturbation when the target flux changes. The VIM speed, indirectly
related to the separation between the companions and their relative brightness, varies between the events in a wide range,
displaying a nonlinear dependence on the amplitude. It seems that the larger photometric dips cause relatively weaker
astrometric excursions. The most prominent VIM companion is outside the range of quasi-linear blending of signals,
i.e., it is probably closer to the boundary of the digital aperture, or, on the contrary, very close to the target star.
Some powerful astrometric perturbations barely noticeable in the light curves are also present in the data.

\section{Periodogram analysis}
\label{per.sec}
\citet{boy} discuss a significant periodical signal detected in the light curve of \st with a period of 0.88 d. 
They do not mention that this periodicity is also present in the astrometric positions of the target. It is obvious,
for example, in Fig. \ref{q8pixf.fig} outside of the main event, especially in the row coordinate (right plot).
The shape of the periodic variations is sharp-crested, which simply reflects the shape of light curve variations.
The 0.88-day flux periodicity is relatively weak but still confidently detectable in our Lomb's $\chi^2$
periodogram analysis, as we will see in the following.

We use the algorithm described by \citet{mak12}, developed for exoplanet search in astrometric and radial velocity
data. The algorithm is based on simple and mathematically rigorous principles, incorporating elements of robust
statistical analysis. In this application, we are using only the first part of the algorithm, namely, the
periodogram analysis. A periodogram is a function $A(P)$ where $P$ is the fitting sinusoid period for a given time
series of data ${\bf d}$ (not necessarily equally spaced) measured at times ${\bf t}$, and $A$ is the amplitude of the
fitting sinusoid. The method proposed by \citet{lom} is to solve the system of linear equations
\eb
[{\bf 1}\: \cos(2\pi{\bf t}/P)\: \sin(2\pi{\bf t}/P)]\; {\bf a}={\bf d}
\ee
by the LS method (possibly introducing weights for measurements of unequal precision) for each trial period $P$.
The sought amplitude is then $A=\sqrt{a_2^2+a_3^2}$. Alternatively, the objective function can be the $\chi^2$
of post-fit residuals. The amplitude periodogram (with peaks indicating significant periodic signals) and the $\chi^2$
periodogram (with dips indicating significant periodic signals) are practically identical as far as the amount of information is concerned, but the latter is more amenable to the confidence (or false alarm probability) estimation
by the standard $F$-test. 

We applied this periodogram analysis to the PCA-filtered flux and astrometric time series of \st for each quarter
of the mission. The purpose was to identify significant periodic signals. In the range of periods between 5 and 90
days, the light curves often include significant periodicities with amplitudes of 100 to 200 ppm. However,
these signals appear to be transient, as neither phase nor period is stable between the quarters. For example,
the highest peak in the $A(P)$ curve rising up to 107 ppm is found at $P=19.7$ d for Q11, but in the next quarter Q12, the
most prominent period is 44 d (although a slightly smaller peak is present at $P=18.5$ d). Such shifting, unstable periodicity is typical for normal, magnetically inactive
and slowly rotating stars, such as the Sun. However, this surmise contradicts the high projected velocity of rotation
determined by \citet{boy}. The possibility that the transient periodicities found in this range are artifacts
caused by the PCA pre-whitening can not be ruled out at this stage. Indeed, some of the principal components
shown in Fig. \ref{5pc.fig}, look sinusoid-like. As a way to resolve this problem, one could
perform a separate periodogram analysis of each of the 4 principle components for each quarter and take into account
the corresponding coefficients of the Tukey bisquare fit. We have not performed this analysis because the results
seem to be negative in any case.

At periods shorter than 5 d, the dominating periodic signal invariably comes at $P=0.88$ d. This periodicity was already
detected and discussed by \citet{boy}, who interpreted it as rotation of the target star. Our calculations with
PCA pre-whitened light curves confirm the presence of this signal. For individual
quarters, the amplitude of the main sinusoidal term is 50 -- 218 ppm. The signal is definitely non-sinusoidal,
which is betrayed by a series of harmonics at one half, one third, etc., the main period. Visually, the 0.88-day
variation in flux is sharp-crested, rather than flat, at the bottom, and resembles the light curves of short-period
eclipsing binaries or rotating ellipsoidal stars. It does not look like a small-amplitude variation caused by transient
and short-lived spots on the rotating surface \citep{mak09}. 

\begin{figure}[htbp]
\epsscale{0.55}
\plotone{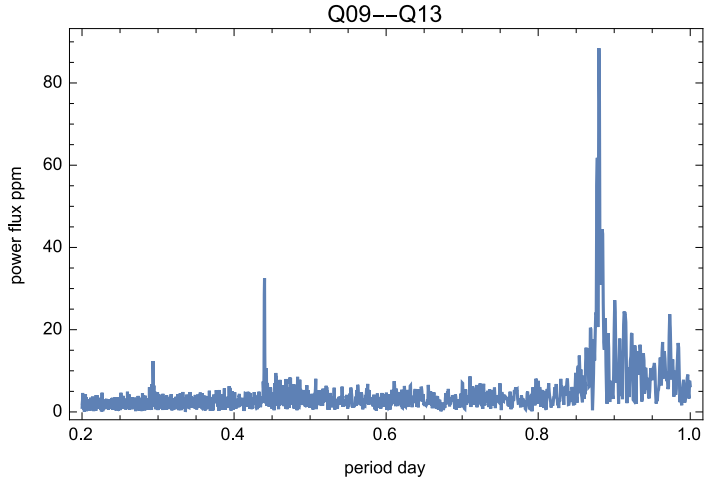}
\hspace{2pc}
\caption{Lomb's periodogram of the combined light curve for \st over 6 quarters, Q9 -- Q15. The strong
peak at $P=0.88$ d and its harmonics (not shown in this graph) indicates that the phase of this periodic variation was stable
over at least 1.5 years of continuous observation.}
\label{perio.fig}
\end{figure}

On the Sun, photospheric spots, plages, and other magnetic features appear with the same probability at any longitude
and very seldom last for longer than one revolution period. The phase of rotation-induced variations in the solar
irradiance curve is therefore scrambled on timescales longer than one month. Even the most magnetically active
stars with much greater amplitudes of variability do not keep the same phase of variation longer than a few months,
unless their activity is enhanced by a close binary companion. On the other hand, F-type stars have shorter periods
of rotation and more persistent photometric features. \citet{mat} investigated the light curves of 22 F-type stars
with periods between 2 and 12 days and found a group of 5 stars with modulations which remained coherent in
phase over the entire duration of observation. They interpreted these stars as rotators with very stable active
longitudes where groups of starspots predominantly emerge. \citet{mat} discarded the possibility of stable classical
pulsations because the stars are not residing in the known instability zones. The morphology of the 0.88 d variations
in the light curve, which are flat at the top and sharp at the bottom, also favor the rotational modulation interpretation.

Non-radial pulsations caused by interior acoustic waves are unlikely to keep a permanent phase for long durations of time.
As a way to probe for the origin of the 0.88-day signal, we computed a periodogram for a combined light curve over
five quarters, Q9 through Q13. This stretch represents a quiescent interval without major photometric peculiarities.
The resulting amplitude periodogram is shown in Fig. \ref{perio.fig}. The 0.88-day peak is very sharp and strong, as 
well as the second and third harmonics of the main mode. If the phase of this variation changed during this 15-month
interval, the peak would be diluted or completely gone, because we fit a single pair of sine and cosine functions to
the entire data set. Thus, the periodic signal is remarkably persistent over 15 months at least.

A significant fraction of field F-type stars possess elevated degrees of coronal magnetic activity, as betrayed by
intense X-ray radiation \citep{suc}. Those active F stars include young stars, binaries of RS CVn and BY Dra-type,
and old, slightly evolved  stars of unclear nature. Young stars of pre-main sequence age and components of tight
binaries are known to rotate fast, which is the reason for their elevated X-ray activity. Our object of investigation,
however, is neither young or binary, nor is it associated with a detectable X-ray source. Here we will argue that this signal comes from one of the blended companions
rather than from KIC 8462852.

\begin{figure}[htbp]
\epsscale{0.55}
\plotone{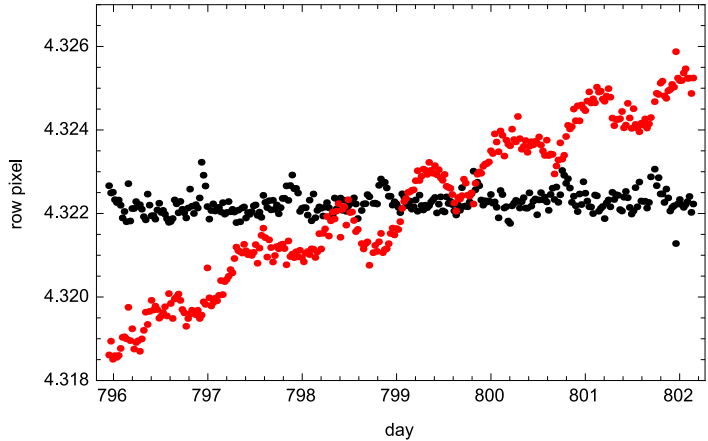}
\hspace{2pc}
\caption{Original (unprocessed) SAP data collected during Q8 of the mission. Black dots represent individual
row (2nd coordinate) pixel coordinates, red dots show the flux measurements rescaled to fit the same range of values
as the coordinates.}
\label{5pix.fig}
\end{figure}

We estimated a VIM speed of $10^{-8}$ -- $10^{-7}$ pix/(e s$^{-1}$) for the major photometric events, which definitely
happened on the target star simply because of the large magnitude of variation (Sect. \ref{eve.sec}). The amplitude of 
the 0.88-day signal is 90 ppm (Fig. \ref{perio.fig}), which is approximately 22 e s$^{-1}$. The expected amplitude
of a VIM in the astrometric trajectory is then of order $10^{-6}$ -- $10^{-5}$ pix. This is too small to be detected
in the Kepler data. However, the 0.88-day periodicity is clearly present in the astrometric data, and {\it sometimes},
its magnitude is quite large, for example, see the right plot in Fig. \ref{q8pixf.fig} for Q8 outside of the major dip
on day 793, where its amplitude is up to $10^{-3}$ pix. Thus, the VIM caused by the 0.88-day periodicity is 100 -- 1000
times stronger in VIM speed than the main event. Furthermore, as Fig. \ref{5pix.fig} shows, the VIM in row pixel
coordinates is counter to the flux variation, whereas they are aligned for the dip. The VIM direction is almost opposite
to the excursion caused by the dip. We detect two distinct VIM events simultaneously perturbing the data,
which originate from different sources. 

\section{A swarm of comets or interstellar junk?}
\label{junk.sec}
\citet{bod} argued that a single comet can not produce a dip in the light curve as deep and as long as the two
major events observed in Q8 and Q16. They propose that a swarm of at least 30 comets traveling in a tight pack can
explain the events during the two last quarters, Q16 and Q17. The dip in Q8 can not be explained by the same swarm
of objects, however. This interpretation invokes multiple groups of comets in significantly different orbits around
the star, speculatively indicating the start of a Late Heavy Bombardment episode. 

We propose a different interpretation which is consistent with the observed peculiarities of the Kepler data and
other information about the star. We consider a large swarm of interstellar objects ranging in size from
comets to planets unrelated to the target star, traveling in the interstellar space, which happened to cross the line
of site to the target and, perhaps, its near neighbors on the sky. The irregularly spaced events are explained as
randomly timed occultations from different parts of the swarm. Such free-traveling swarms can be the remnants of
catastrophic disintegration of a rich exoplanetary system or a star-formation episode in a depleted molecular cloud.
The interstellar Na D absorption lines detected by \citet{boy} are likely to be related to the foreground cloud.
A similar hypothesis was recently discussed by \citet{wrs}. Alternatively, a swarm of comets and debris orbiting another foreground star 
which accidentally happened to be close to the
target star in the sky projection (an optical pair) can also be considered.

The existence of interstellar comets has been suspected for some time but no direct evidence has yet been found.
Passing through the inner part of the Solar system, they would be detectable as comets on high-eccentricity hyperbolic
orbits. \citet{sen} were first to estimate the rate of such encounters and concluded that they should be very rare.
A more recent estimation \citep{coo} suggests that fast-moving interstellar comets can realistically be detected by
the LSST during its 10-year nominal mission. All these estimates assume randomly uniform distribution of comets in space,
whereas in reality, they may be grouped in swarms.

What kind of density should a swarm of comet-like or planetoid-like objects have to produce the observed rate of
irregular occultations? Assuming a radius of $1.7\,R_\sun$ for \st \citep{hu} and a distance of 500 pc to the
star, the cross-section of the
occultation events during the 4 years of Kepler main mission is $0.125\times 10^{-6}\mu$ in arcsec$^2$, where
$\mu$ is the relative proper motion of the swarm and the target star in mas yr$^{-1}$. The expected number of
occultation events is, approximately,
\eb
N\approx 10^{-3}\left[\frac{5 D_j}{D_s}\right]^2\left[\frac{\rho}{{\rm AU}^{-2}}\right]\left[\frac{\mu}{1\:{\rm mas}
\: {\rm  yr}^{-2}}
\right]
\ee
where $D_s$ is the distance to the star, $D_j$ is the distance to the foreground swarm, and $\rho$ is the surface number density
of objects. Since the relative
proper motion is expected to be of order 10 mas yr$^{-1}$, a surface density of 1000 per AU$^2$ is sufficient
to produce the observed number of events at a distance ratio of 5, which is further relaxed to $\sim 40$ if the
swarm is close to the target star. Such surface densities are probably feasible, considering, for example, the approximate
volume density $10^5$ AU$^{-3}$ of asteroids larger than 1 km in diameter in the Solar asteroid belt. However, the sizes of
these objects should be much greater than anything we observe in the Solar system. An object producing a
20\% dip in flux by direct occultation at a distance ratio of 1/5 should be comparable in size to Jupiter. A more likely alternative
invokes comets shrouded in dense cocoons of dust and debris. Speculatively, collisional break-up of tightly packed planetary systems
can result in debris clouds with such peculiar properties.
\label{con.sec}

The relative proper motion can be estimated assuming a distance of 500 pc and a typical duration of occultation of
1 d. The upper limit is 6 mas yr$^{-1}$, which is a small number considering that the proper motion of the star
itself is greater than 10 mas yr$^{-1}$. The distance estimation for \st is quite uncertain, however. The Gaia DR1 data
\citep{bra} provides a trigonometric distance of 392 ($+54,-42$) pc, but the actual uncertainty is higher
because of the unknown zero-point bias which may come up to 0.3 mas in parallax. The smaller distance is more consistent
with the colors determined in the AAVSO photometric survey \citep{he}, $B-V=0.508\pm0.062$, $g'-r'=0.35\pm0.06$ mag.
A smaller distance to the star implies a higher relative proper motion.

\begin{figure}[htbp]
\epsscale{0.95}
\plotone{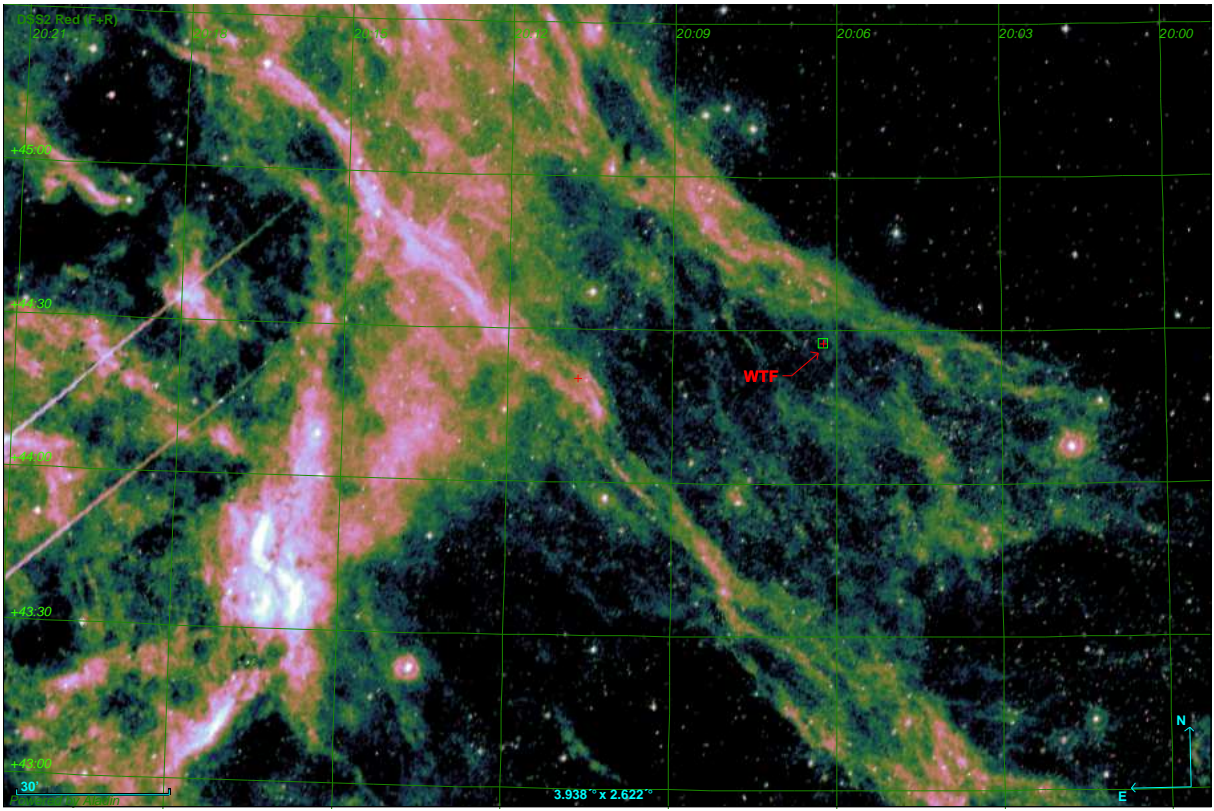}
\hspace{2pc}
\caption{A map of the extended area including the star in question \st marked with a small green square
labeled ``WTF", the peculiarly shaped Simeiz 57
nebula (also known as the Propeller Nebula) in the lower left corner, and surrounding H$_\alpha$-emission filaments,
rendered from the DSS2-red images using the Aladin application from CDS. The false color map is adjusted to emphasize the
structure of the low-density gaseous filaments.}
\label{map.fig}
\end{figure}

As a clue in support of the foreground swarm hypothesis, we find that \st is located at the southwest tip of the extended
filament-like H$_\alpha$-emission nebula DWB 123 \citep{dic}. This optically visible HII cloud with dimensions $74\farcm 4$
by $12.\farcm2$ extends northeast from the star's position by more than $1\degr$ in an almost linear shape. Originally thought to be
associated with the distant Cygnus X complex in the Orion arm, the cloud is more likely related to the web of filaments
around the peculiar-shaped emission-line nebula Simeiz 57 (often spelled Simeis 57 in the literature), discovered by Soviet astronomers
in the Crimean Astrophysical Observatory \citep{gaz}. Still very little is known about this nebula, but it may be nearby
\citep{isr}. Fig. \ref{map.fig} shows a map of the extended area including Simeiz 57 (also known as the Propeller Nebula, in the
southeast corner), the filamentary emission clouds adjoining it, and the location of KIC 8462852 marked as WTF\footnote{WTF is
a popular nickname for the star derived from the title of the discovery's paper ``Where is the flux?" \citep{boy}.}. 
Neither distance nor illuminating star are known for DWB 123 but the large extent and low electron density suggests
its proximity. It would be interesting to find the illuminating star and to check if this cloud is closer to us than \st.

A cloud of foreground comets shrouded in dense dust, 
whether solivagant (traveling by themselves) or attached to a foreground star, analogous to our
system's Kuiper belt, could explain the irregular arrival of occulters, the varying depth and duration of observed dips, and
the magnitude of these events. A long-term photometric monitoring of \st may reveal more signs of foreground obscuration by dust.

\section{Summary}
\label{sum.sec}
The complex pattern of photometric variability detected for \st is often, but not always, faithfully reflected in
correlated astrometric excursions called VIM. This confirms a non-negligible amount of blending in the detected
aperture flux. While most of the prominent dips in the light curve and the corresponding VIM events indicate
a single direction to the largest perturbing component (PA$\approx 330\degr$), proving that they occur on the
intended target, the magnitude of astrometric perturbations shows a large scatter and a nonlinear dependence on
the photometric amplitude. A very close unresolved companion is possible. An alternative explanation is a distant
($\gtrsim 20"$) but brighter companion. A number of VIM events, however, have much greater VIM speeds and different
directions. These are caused by photometric events on the faint optical companions of the target. The small-scale
variability observed outside of the main dips may be caused by the blended signals.

The most remarkable 0.88 d variability previously attributed to KIC 8462852, appears to be coming from a contaminant, but not
from the same companion as the most pronounced VIMs. Per Eq. \ref{vim.eq}, the observed VIM speed ($ds/df$)
is very sensitive to the origin of photometric variability when two star images are blended. If the variability occurs on
the faint companion, the VIM speed is much greater and the VIM direction is roughly opposite to the main events.
The phase (but not the observed amplitude) of this non-sinusoidal wave is found to be remarkably stable over at least 15 months of continuous data. The possible origin of this signal is
synchronous rotation of an optical companion, unrelated to the target star. Transient periodicity occurs in both
light curve and astrometry at longer periods of 10 -- 40 d, but it is not persistent in phase, amplitude, or
frequency.

The observed pattern of stochastic, randomly spaced photometric events and the complex character of VIM interference
point at a foreground swarm of comet-like objects or planets as a more plausible interpretation than a
family of highly eccentric comets orbiting the target star. The swarm may be a free-traveling interstellar group
of objects or a belt associated with the hypothetical optical companion.
\section*{Acknowledgments}
Much of the data presented in this paper were obtained from the Mikulski Archive for Space Telescopes (MAST). STScI is operated by 
the Association of Universities for Research in Astronomy, Inc., under NASA contract NAS5-26555. Support for MAST for non-HST data 
is provided by the NASA Office of Space Science via grant NNX09AF08G and by other grants and contracts.
This research has made use of ``Aladin sky atlas" developed at CDS, Strasbourg Observatory, France. 
The Digitized Sky Surveys used in this paper were produced at the Space Telescope Science Institute under U.S. Government grant 
NAG W-2166. The images of these surveys are based on photographic data obtained using the Oschin Schmidt Telescope on Palomar Mountain 
and the UK Schmidt Telescope. This publication makes use of data products from the Two Micron All Sky Survey, which is a joint 
project of the University of Massachusetts and the Infrared Processing and Analysis Center/California Institute of Technology, 
funded by the National Aeronautics and Space Administration and the National Science Foundation. This work has made use of data from the 
European Space Agency (ESA)
mission {\it Gaia} (\url{http://www.cosmos.esa.int/gaia}), processed by
the {\it Gaia} Data Processing and Analysis Consortium (DPAC,
\url{http://www.cosmos.esa.int/web/gaia/dpac/consortium}). Funding
for the DPAC has been provided by national institutions, in particular
the institutions participating in the {\it Gaia} Multilateral Agreement.

\label{lastpage}

\end{document}